# Reduction in iron losses In Indirect Vector-Controlled IM Drive Using FLC

C. Srisailam  
Electrical Engineering Department  
Jabalpur Engineering College,  
Jabalpur, Madhya Pradesh  
Email:chikondra007@gmail.com

Mukesh Tiwari  
Electrical Engineering Department  
Jabalpur Engineering College  
Jabalpur, Madhya Pradesh  
Email:mukesh_tiwari836@yahoo.co.in

Dr. Anurag Trivedi  
Electrical Engineering Department  
Jabalpur Engineering College  
Jabalpur, Madhya Pradesh  
Email:dr.anuragtrivedi@yahoo.co.in

*Abstract-* **This paper describes the use of fuzzy logic controller for efficiency optimization control of a drive while keeping good dynamic response. At steady-state light-load condition, the fuzzy controller adaptively adjusts the excitation current with respect to the torque current to give the minimum total copper and iron loss. The measured input power such that, for a given load torque and speed, the drive settles down to the minimum input power, i.e., operates at maximum efficiency. The low-frequency pulsating torque due to decrementation of flux is compensated in a feed forward manner. If the load torque or speed commands changes, the efficiency search algorithm is abandoned and the rated flux is established to get the best dynamic response. The drive system with the proposed efficiency optimization controller has been simulated with lossy models of converter and machine, and its performance has been thoroughly investigated.**

*Key words: Fuzzy logic controller (FLC), Fuzzy logic motor control (FLMC), Adjustable speed drives (ASDs).*

## I. INTRODUCTION

Efficiency improvement in variable frequency drives has been getting a lot of attention in recent years. Higher efficiency is important not only from the viewpoints of energy saving and cooling system operation, but also from the broad perspective of environmental pollution control. In fact, as the use of variable speed drives continues to increase in areas traditionally dominated by constant speed drives, the financial and environmental payoffs reach new importance.

A drive system normally operating at rated flux gives the best transient response. However, at light loads, rated flux operation causes excessive core loss, thus impairing the efficiency of the drive. Since drives operate at light load most of the time, optimum efficiency can be obtained by programming the flux. A number of methods for efficiency improvement through flux control have been proposed in the literature. They can be classified into three basic types. The simple precompiled flux program as a function of torque is widely used for light load efficiency improvement. The second approach consists in the real time computation of losses and corresponding selection of flux level that results in minimum losses. As the loss computation is based on a machine model, parameter variations caused by temperature and saturation effects tend to yield suboptimal efficiency operation. The on-line efficiency optimization control [1]-[3] on the basis of search, where the flux is decremented in steps until the measured input power settles down to the lowest value, is very attractive. The control does not require the knowledge of machine parameters, is completely insensitive to parameter changes, and the algorithm is applicable universally to any arbitrary machine.

## II. CONTROL SYSTEM DESCRIPTION

"Fig. 1", show the block diagram of an indirect vector controlled induction motor drive incorporating the proposed efficiency optimization controller. The feedback speed control loop generates the active or torque current command ($i_{qs}^*$) as indicated. The vector rotator receives the torque and excitation current commands $i_{qs}^e$ and $i_{ds}^e$ respectively, from the two positions of a switch: the transient position (l), where the excitation current is established to the rated value ($i_{ds,r}$) and the speed loop feeds the torque current and the steady state position, where the excitation and torque currents are generated by the fuzzy efficiency controller and feed forward torque compensator which will be explained later.

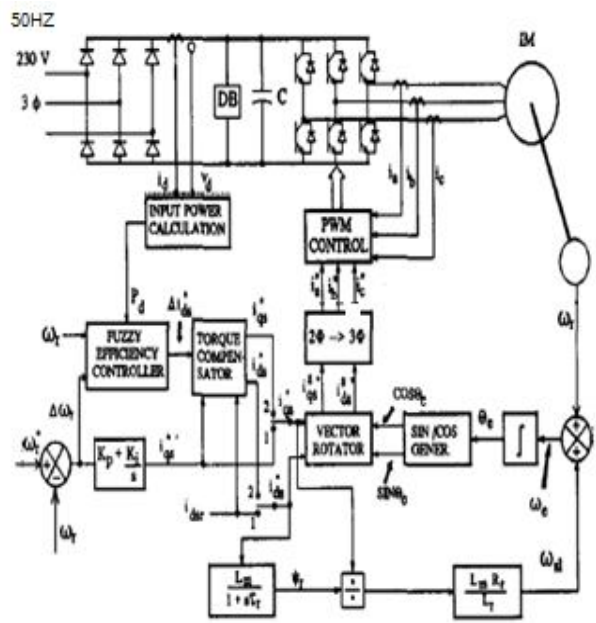

Fig.1. Efficiency optimization controller. Indirect vector controlled induction motor drive incorporating the FLC.

The fuzzy controller becomes effective at steady-state condition, i.e., when the speed loop error $\Delta\omega_r$, approaches zero. Note that the dc link power $P_d$, instead of input power, has been considered for the fuzzy controller since both follow symmetrical profiles. The principle of efficiency optimization control with rotor flux programming at a steady-state torque and speed condition is explained in "Fig. 2". The rotor flux is decreased by reducing the magnetizing current, which ultimately results in a corresponding increase in the torque current (normally by action of the speed controller), such that the developed torque remains constant. As the flux is decreased, the iron loss decreases with the attendant increase of copper loss. However, the total system (converter and machine) loss decreases, resulting in a decrease of dc link power. The search is continued until the system settles down at the minimum input power point A, as indicated. Any excursion beyond the point A will force the controller to return to the minimum power point.

*A. Efficiency Optimization Control*

"Fig.2" explains the fuzzy efficiency controller operation. The input dc power is sampled and compared with the previous value to determine the increment $\Delta P_d$. In addition, the last excitation current decrement (L$\Delta i_{ds}$) is reviewed. On these bases, the decrement step of $\Delta i_{ds}^*$ is generated from fuzzy rules through fuzzy inference and defuzzification [4], as indicated. The adjustable gains $P_b$ and $I_b$, generated by scaling factors computation block, convert the input variable and control variable, respectively, to per unit values so that a single fuzzy rule base can be used for any torque and speed condition. The input gain $P_b$ as a function of machine speed $\omega_r$ can be given as

$$P_b = a\omega_r + b \quad (1)$$
$$I_b = c_1\omega_r - c_2\tilde{T}_e + c_3 \quad (2)$$

Where $\tilde{T}_e = K_t' i_{ds}^* i_{qs}^* \quad (3)$

where the coefficients $a \& b$ was derived from simulation studies. The output gain $I_b$ is computed from the machine speed and an approximate estimate of machine torque ($T_e$).

Efficiency Optimization Control Fig.3 explains the fuzzy efficiency controller operation. The input dc power is sampled and compared with the previous value to determine the increment $\Delta P_d$. In addition, the last excitation current decrement (L$\Delta i_{ds}$) is reviewed. On these bases, the decrement step of $\Delta i_{ds}^*$ is generated from fuzzy rules through fuzzy inference and defuzzification [4], as indicated. The adjustable gains $P_b$ and $I_b$, generated by scaling factors computation block, convert the input variable and control variable, respectively, to per unit values so that a single fuzzy rule base can be used for any torque and speed condition. The input gain $P_b$ as a function of machine speed $\omega_r$ can be given as where the coefficients $a \& b$ was derived from simulation studies. The output gain $I_b$ is computed from the machine speed and an approximate estimate of machine torque ($T_e$). The appropriate coefficients $c_1, c_2$ and $c_3$ were derived from simulation studies. A few words on the importance of the input and output gains are appropriate here. In the absence of input and output gains, the efficiency optimization controller would react equally to a specific value of $\Delta P_d$, resulting from a past action $\Delta i_{ds}^*(k-1)$, irrespective of operating speed. Since the optimal efficiency point A (see Fig. 2) is speed dependant, the control action could easily be too conservative, resulting in slow convergence, or excessive, yielding an overshoot in the search process with possible adverse impact on system stability. As both input and output gains are function of speed, this problem does not arise. Equation (2) also incorporates that a priori knowledge that the optimum value of $i_{ds}^*$ is a function of torque as well as machine speed. In this way, for different speed and torque conditions, the same $\Delta i_{ds}^*$ (pu) will result in different $\Delta i_{ds}^*$, ensuring a fast convergence. One additional advantage of per unit basis operation is that the same fuzzy controller can be applied to any arbitrary machine, by simply changing the coefficients of input and output gains.

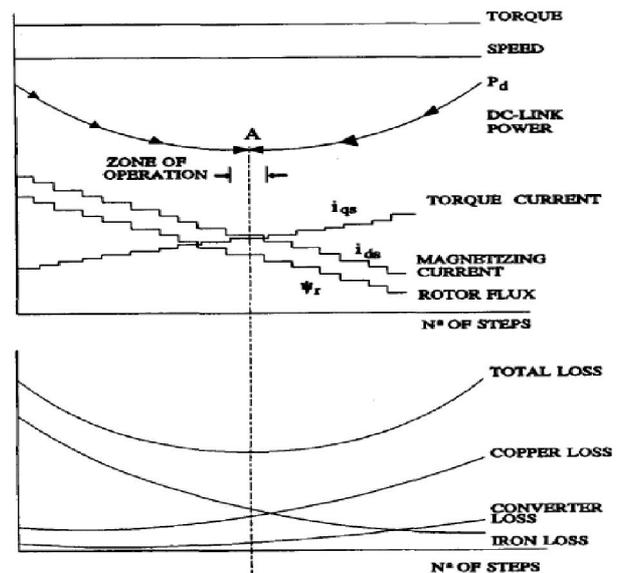

Fig.2. Principle of efficiency optimization control with rotor flux programming

The membership functions for the fuzzy efficiency controller are shown in Fig. 4. Due to the use of input and output gains, the universe of discourse for all variables are normalized in the [-1, 1] interval. It was verified that, while the control variable $\Delta i_{ds}^*$, required seven fuzzy sets to provide good control sensitivity, the past control action $L\Delta i_{ds}^*$ (i.e., $\Delta i_{ds}^*(k - 1)$) needed only two fuzzy sets, since the main information conveyed by them is the sign. The small overlap of the positive (P) and negative (N) membership functions is required to ensure proper operation of the height defuzzification method [4], i.e., to prevent indeterminate result in case $L\Delta i_{ds}^*$ approaches zero.

An example of a fuzzy rule can be given as IF the power increment ($\Delta P_d$) is negative medium (NM) and the last $\Delta i_{ds}^*$ ($L\Delta I_{dS}$) is negative (N), THEN the new excitation increment ($\Delta i_{ds}^*$) is negative medium (NM). The basic idea is that if the last control action indicated a decrease of dc link power, proceed searching in the same direction, and the control magnitude should be somewhat proportional to the measured dc link power change. In case the last control action resulted in an increase of $P_d$ ($\Delta P_d > 0$), the search direction is reversed, and the $\Delta i_{ds}^*$ step size is reduced to attenuate oscillations in the search process.

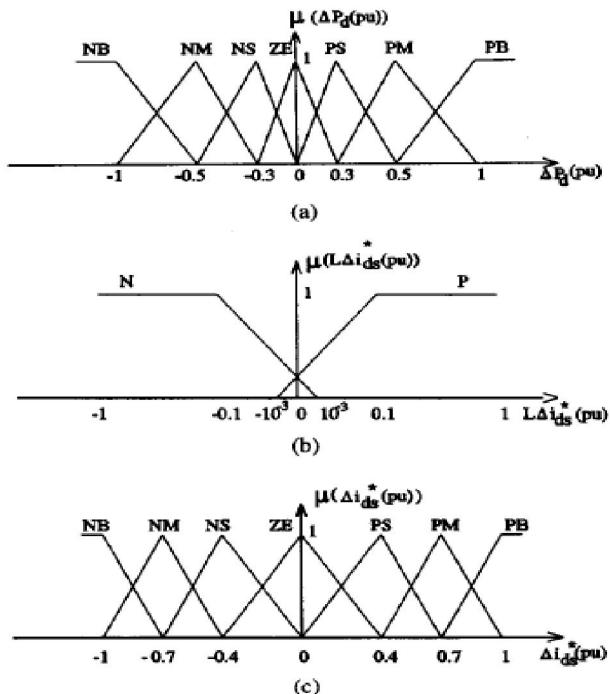

Fig.4. Membership functions for efficiency controller. (a) Change of dc link power $\Delta P_d$ (pu). (b) Last change in excitation current ($L\Delta i_{ds}^*$(pu)). (c) Excitation current control increment ($\Delta i_{ds}^*$ (pu)).

### B. Feed forward Pulsating Torque Compensation

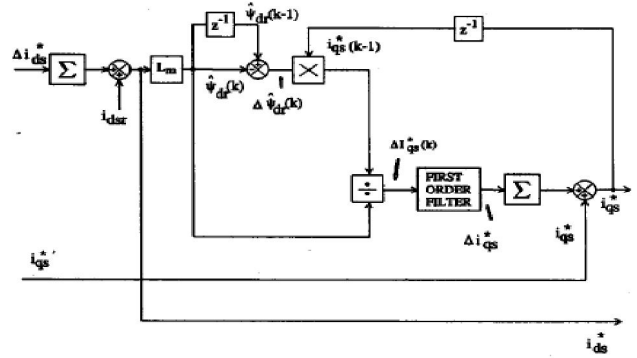

Fig. 5. Feed forward pulsating torque compensator block diagram

As the excitation current is decremented with adaptive step size by the fuzzy controller, the rotor flux $\Psi_{dr}$ will decrease exponentially [5], which is given by:

$$\frac{d}{dt}\Psi_{dr} = \frac{L_m i_{ds} - \Psi_{dr}}{\tau_r} \quad (4)$$

The decrease of flux causes loss of torque, which normally is compensated slowly by the speed control loop. Such pulsating torque at low frequency is very undesirable because it causes speed ripple and may create mechanical resonance. To prevent these problems, a feed forward pulsating torque compensator has been proposed. Under correct field orientation control, the developed torque is given by

$$T_e = K_t i_{qs} \Psi_{dr} \quad (5)$$

For an invariant torque, the torque current $i_{qs}$, should be controlled to vary inversely with the rotor flux. This can be accomplished by adding a compensating signal $\Delta i_{qs}^*(t)$ to the original $i_{qs}^*$ to counteract the decrease in flux $\Delta \Psi_{dr}(t)$, where $t \in [O,T]$ and T is the sampling period for efficiency optimization control. Let $i_{qs}(0)$ and $\Psi_{dr}(0)$ be the initial values for $i_{qs}$ and $\Psi_{dr}$, respectively, for the k-th step change of $i_{ds}^*$. For a perfect compensation, the developed torque must remain constant, and the following equality holds:

$$\Psi_{dr}(0) + \Delta \Psi_{dr}(t)][i_{qs}(0) + \Delta i_{qs}(t)] = \Psi_{dr} i_{qs}(0)$$

Solving for $\Delta i_{qs}(t)$ yields $\quad (6)$

$$\Delta i_{qs}(t) = \frac{-\Delta \Psi_{dr}(t) i_{qs}(0)}{\Psi_{dr}(0) + \Delta \Psi_{dr}(t)} \quad (7)$$

Eq (7) is adapted to produce $\Delta i_{qs}^*(t)$

Compensated torque current step is computed by

$$\Delta I_{qs}^*(k) = \frac{\Psi_{dr}(k-1) - \Psi_{dr}(k)}{\Psi_{dr}(k)} i_{qs}^*(k-1) \quad (8)$$

## III. RESULTS AND DISCUSSION

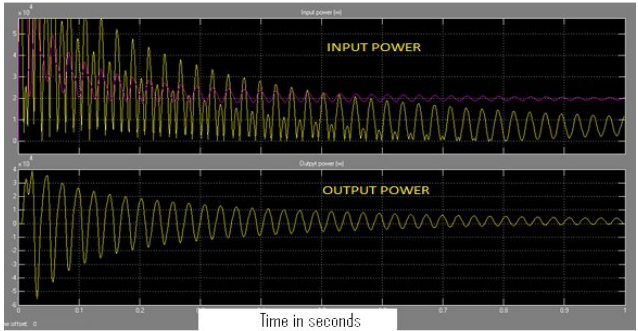

Fig.6 Input and output power when FLC is not used

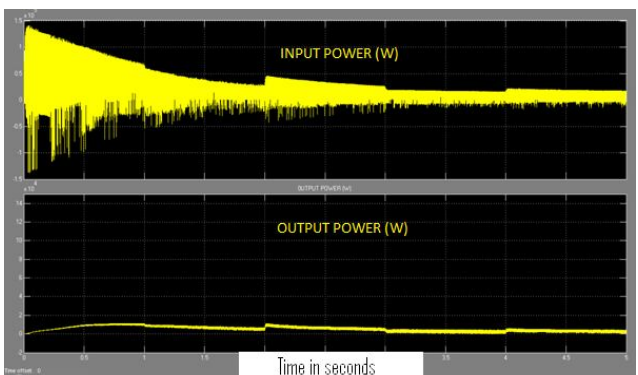

Fig.7 Input and output power when FLC used

TABLE.I: When FLC is not used with drive

| Load torque(N-m) w.r.t. F.L | Input power(kw) | Output power(kw) | Efficiency (%) |
|---|---|---|---|
| 6(1/4th F.L) | 8.4 | 0.98 | 11 |
| 8(1/3rd F.L) | 7.9 | 1.2 | 16 |
| 12(1/2 F.L) | 6.9 | 1.88 | 26 |
| 18(3/4th F.L) | 5.5 | 2.8 | 50 |

TABLE.II: When FLC is used with drive

| Load torque(N-m) w.r.t. | Input power(kw) | Output power(kw) | Efficiency (%) |
|---|---|---|---|
| 6(1/4th F.L) | 2.4 | 1 | 41 |
| 8(1/3rd F.L) | 2.5 | 1.2 | 50 |
| 12(1/2 F.L) | 3.5 | 1.9 | 57 |
| 18(3/4th F.L) | 4.5 | 2.7 | 60 |

## IV. CONCLUTION

Adjustable speed drives which allow the control of speed of rotation of induction motor provide significant savings in energy requirements of motor when operating at reduced speeds and torques. This was observed mainly through minimize the input power at any speed and torque. The objective of this paper is to study fuzzy controls which optimize the adjustable speed drives on the basis of energy efficiency. It was observed that efficiency has been optimized by minimizing the input power and the output power is maintained constant with the help of Fuzzy controller. For a given power output of the induction motor; its input power is supplied as input to the fuzzy controller. Constant power output and a reduced input the graphs obtained as outputs from the MATLAB simulink.

ACKNOWLEDGMENT

We take this great opportunity to convey our sincere thanks to CERA 2009 organising committee for recognising our efforts by selecting our paper for oral presentation. We also extend our sincere gratitude to our head of department who gave inspiring suggestions while preparing this paper.

## Author's profile

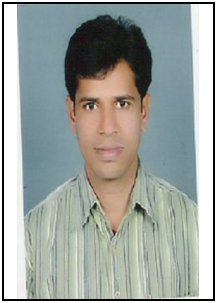

*Mr. risailam* was born in Andhradesh Pradesh at Mahaboob nagar on 15$^{th}$ may 1979.received B.E degree in (Electrical&Electronics ) from Vasavi college of Engireeringin 2006 now is the student of master of Engineering in (Control system) Department of Electrical Engineering. Jabalpur Engineering College Jabalpur (M.P) INDIA. His research interest on fuzzy logic, drives and control system. had published one International conference& twelve National Conferences.

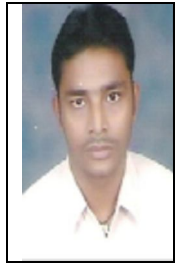

*Mr. mukesh tiwari* was born in Madhya-Pradesh at katni on 27$_{th}$ November 1983. He received B.E degree in (Electronic & Communication) from Rewa Institute of Technology Rewa in 2007 he is the student of master of Engineering in (Control system) Department of Electrical Engineering. Jabalpur Engineering College Jabalpur (M.P) INDIA. His research interest on fuzzy logic, communication, and control system He has published two International journal & Three National Conferences

*Dr.Anurag Trivedi* received B.E. degree in Electrical Engg. from Jabalpur Engg. College, Jabalpur in 1987 and the Ph.D. degree in power systems Indian Institute of Technology, Roorkee in 2006. Is currently Reader in Electrical Engg. department. his research interest on power system fault analysis,Microprocessors and power drives. Had published one International conference& eighteen National Conferences.